\documentclass[acmlarge]{acmart}
\setcopyright{acmcopyright}
\copyrightyear{2023}
\acmYear{2023}
\acmDOI{XXXXXXX.XXXXXXX}
\begin{document}
\title
{Dynamic Documentation for AI Systems}

\author{Soham Mehta}
\authornote{Both authors contributed equally to this research.}
\email{sgm2160@columbia.edu }
\author{Anderson Rogers}
\authornotemark[1]
\email{ar745@cornell.edu }
\author{Thomas Krendl Gilbert}
\email{thomaskrendlgilbert@gmail.com}
\renewcommand{\shortauthors}{Mehta, Rogers, and Gilbert}


\begin{abstract}
  AI documentation is a rapidly-growing channel for coordinating the design of AI technologies with policies for transparency and accessibility. Calls to standardize and enact documentation of algorithmic harms and impacts are now commonplace. However, documentation standards for AI remain inchoate, and fail to match the capabilities and social effects of increasingly impactful architectures such as Large Language Models (LLMs). In this paper, we show the limits of present documentation protocols, and argue for \textit{dynamic documentation} as a new paradigm for understanding and evaluating AI systems. We first review canonical approaches to system documentation outside the context of AI, focusing on the complex history of Environmental Impact Statements (EISs). We next compare critical elements of the EIS framework to present challenges with algorithmic documentation, which have inherited the limitations of EISs without incorporating their strengths. These challenges are specifically illustrated through the growing popularity of Model Cards and two case studies of algorithmic impact assessment in China and Canada. Finally, we evaluate more recent proposals, including Reward Reports, as potential components of fully dynamic AI documentation protocols.
\end{abstract}




\maketitle

\section{Introduction}
The Office of Science and Technology Policy (OSTP) recently released the \textit{AI Bill of Rights Framework} to outline recommended best practices for algorithmic transparency to protect Americans from harm \cite{blueprint22}.
Yet the blueprint treats automated decision-making and AI interchangeably, without rigorously defining either term. This is unfortunate, as machine learning practitioners have already taken up documentation tools as a critical interface for the simultaneous design and evaluation of system components. In tension with this development, the failure to distinguish specific policy priorities for AI has limited critical forms of accountability, including the ability to identify which system components are responsible for specific impacts.

Artificial Intelligence began as a field of mathematics based on approximately simulating mechanisms of human cognition through computation \cite{mccarthy2006proposal}, but has evolved to include the design of systems that incorporate multiple types of feedback to operate strategically in deployment environments \cite{gilbert2022choices}. We define AI as “the design of rational agents that select actions to maximize their (expected) utility”\cite{russell2010artificial}. AI is different from the general field of algorithm design, which comprises “[procedures] for solving a mathematical problem in a finite number of steps, often through the repetition of some operation or calculation” \cite{mish2004merriam}. This couples well with the mechanisms involved with automated decision-making, as they both depend on the type and structure of data being used to carry out these decision-making mechanisms. An automated decision system is the “combination of data and learning algorithms to make consequential decisions affecting people or the environment”\cite{harris2005automated}. 

In light of these definitions, the \textit{AI Bill of Rights Framework} is a bit of a misnomer. As written, the Framework articulates policies that may protect Americans from algorithmic and automated decision-making, but not the highly strategic, dynamic, feedback-laden behaviors specific to AI. This paper discusses how documentation protocols must evolve to meet the challenges posed by AI specifically, based on a comparison of past and present documentation techniques both within and beyond the world of AI. Section 2 of this paper outlines the history of Environmental Impact Statements in the United States. Section 3 argues the need for fully dynamic documentation of AI systems, focusing on the limitations of current tools like Model Cards and algorithmic impact assessments as well as more recent proposals like Reward Reports. Section 4 concludes.

\section{Environmental Impact Statements as a Model for Documentation}

Accountability via documentation has long been the goal of public-facing government agencies and organizations. 
In the United States, a crucial example stems from the 1970 National Environmental Policy Act (NEPA). NEPA codified Environmental Impact Statements (EISs), originally intended to document and analyze the social and environmental impacts of engineering and construction projects on populations. These documents and their creation process have been used to better inform policymakers about the “major actions significantly affecting the quality of the human environment”\cite{friesema1976social}. However, the incremental documentation procedure made the promotion of a holistic approach difficult. Due to institutional norms and expectations, policymakers required to evaluate the effects of these practices and policies were only able to do so in a piecemeal manner\cite{friesema1976social}. As a result, these attempts to make decisions with social impacts in mind became justifications for the impacts of projects rather than a mechanism for holding those impacts (and predictions of impacts) accountable. Another aspect of EISs relevant to algorithmic decision-making is the space for public commentary. NEPA required participating agencies to formally respond to every comment they received about the statement, regardless of whether or not the comment aligned within the participating agencies’ scope of work or goals. This had the effect of “opening up agency decision-making processes and counteracting the closed agency pattern…”\cite{friesema1976social}. 

Many aspects of algorithmic decision-making are already beyond the expectations and norms created by EISs. Most notably, policymakers are faced now with counteracting the intangibility of algorithmic harm, as it is caused by a set of opaque, abstract technologies. However, the common thread is opacity of both how systems interact with stakeholder goals and the inability to evaluate cumulative impacts\cite{burdge1990social}.

\subsection{Impact Assessment}

EISs frequently fell victim to indeterminacies of impact assessment, in particular important distinctions between \textit{socio-economic impacts} and \textit{social impacts}. Given the incremental decision-making that is requisite to the norms of government agencies, these terms are often conflated. Socio-economic impacts are given the spotlight, while social impacts are not given their own scope of understanding. Social impacts are harder to define and may lack consensus. For our purposes, a social impact is the "[impact] of a proposed project or policy change on the individuals and social groups within a community or on an entire community"\cite{burdge1990social}. This differs from the broader implications of socio-economic impacts, which may also be lumped in with political impacts. 

The consequence of conflating these impacts is that they are more difficult to assess and be evaluated by a workforce able to appropriately quantify these impacts according to their magnitude and significance. When paired with a lack of planning and proper training, these problems become intractable. Engineers and government planners alike were poorly trained in planning protocol and procedures.\cite{burdge1991brief} The individuals who are hired to oversee these projects are not trained practitioners, and, according to one study, are often used to fill affirmative action quotas.\cite{burdge1991brief} In order for these systems and projects to succeed, they need a qualified and dedicated workforce. 

\subsection{Importance of Public Opinion and Commentary}

Beyond varied impacts, methods of information dissemination have shifted over time. In the history of EISs and similar impact assessments, such as Environmental Impact Assessments (EIAs) and Social Impact Assessments (SIAs), there exist distinct methods of public commentating. Following the codification of EISs in the 1970s, agencies were required by law to respond to and acknowledge all comments left on their projects, regardless of whether or not the comment was relevant to the agencies' motivations.\cite{friesema1976social} In the early years of EISs, these public comments and their responses were listed individually. One example of this can be found in the fifteen pages which comprise section 8 of the EIS detailing the Strategic Petroleum Reserve at Bryan Mound Salt Dome in Brazoria County, TX.\cite{marx1936hoskins} The final EIS was released in December 1977. In more modern EISs, the number of comments on certain projects has increased exponentially. In an EIS from a project in 2020, the executive summary cited receiving and responding to approximately 59,000 comments.\cite{us2020executive} 

This significant increase in public input and comment volume is unsurprising considering the technological advancements made over the past forty years. However, it does pose a difficult set of administrative issues for both participating private agencies and government agencies alike, as was previously mentioned in the prior subsection. The workforce rarely exists to conduct accountability at the scale exhibited by the aforementioned 2020 project.  It begs the question why systems that involve algorithmic decision-making and/or AI, which have social, socio-economic, and political impacts at even greater scales, are not held to the same levels of public accountability, let alone documentation protocols of similar sophistication to EISs. 

\section{The Case for Dynamic Documentation}

The modern incarnation of the Environmental Impact Statement paradigm is AI documentation systems that do not confront the dynamic, feedback-laden nature of machine learning systems. With the revelations of technology industry scandals like the FaceBook Files, it is clear that supposedly “unforeseen” effects of deployed algorithms find their source in intentional algorithmic design choices \cite{WSJ2021}. Studying the design of AI systems with the pitfalls of algorithm design in mind is a step in the right direction. But when AI ethics practitioners fail to contrast the dynamic, utility-maximizing nature of AI with the static and repetitive qualities of automated decision-making, documentation turns into a mere restatement or justification for AI’s negative impacts. As a result, tools that try to identify potential biases or harm using static machine learning (ML) methods, like Model Cards \cite{mitchell2019model} and algorithmic assessments \cite{reisman2018algorithmic}, replicate the flaws of EISs. While valuable, these tools lose sight of the environments in which ML systems are deployed, and do not continuously monitor how AI performs iteration to iteration.

\subsection{The Limits of Model Cards}

Model Cards remain a watershed achievement for transparency in ML. However, this framework would be practically useless in explaining the behavior of a modular and complex system like Meta’s BlenderBot 3 \cite{shuster2022blenderbot}, which combines a number of distinct models and automated systems with interconnected roles in generating responses and monitoring responses for safety. The quintessential case study surrounding Model Cards conducted by Mitchell et. al tests the performance of a smiling classifier on the public CelebA dataset, noting false positives, false negatives, and omissions to detect for age, gender, and race-based biases. This assessment approach assumes that a snapshot of model performance taken at one time can inform general conclusions about how a system meets certain ethical desiderata. While this may be a fair assumption for some automated systems, conducting a Model Card of every classifier in BlenderBot 3 would not generate any actionable insights, as the performance of a single model is dependent on a variety of other systems and feedback. For example, the chatbot’s adversarial/non-adversarial classifier is predicated upon dynamic user feedback, pre-deployment crowdworker evaluations of the system, and a set of safety classifiers called SafetyMix that is used to process crowdworker feedback. Applying a granular and static approach like Model Cards to an ML system like BlenderBot3 would not help designers or policymakers comprehend system failures and pinpoint recommendations. In other words, for complex ML systems, Model Cards are unable to see the forest for the trees.

\subsection{The Limits of Algorithmic Impact Assessments}

On the other hand, algorithmic impact assessments, such as those used in Canada and China, generally are preoccupied with the effects of a system rather than its technical details. In this way, they provide an abstracted account of what is going on “under the hood”. If we examine a completed algorithmic assessment of a tool that reads and assesses employee reviews in Canada, there is little discussion of the reasons that this system might produce undesirable effects. It begins with a seemingly subjective series of yes/no questions with no opportunity for elaboration including “Are the stakes of the decision very high?” and “The algorithmic process will be difficult to interpret or explain” \cite{gilbert2022reward}. While the questionnaire briefly inquires about training data, it only goes as far as asking about the general type of data. The mitigation questions merely ask the designers whether there are documented processes in place to handle unexpected impacts, again failing to require AI practitioners to disclose design details that may eventually produce these unwanted consequences. Since there is neither any documentation robustly logging how the designers of the system believe it would operate or any detail on the training, design elements, or feedback that govern the behavior of this system, Canada’s documentation system does not provide policymakers and stakeholders the requisite information needed to hold the designers liable for the effects of their system or make substantive recommendations to improve the system. It may well function primarily as a bureaucratic rubber stamp, or an extra hoop to jump through to publicly frame technologies as safe and vetted, even when no meaningful examination has actually occurred.

China’s Internet Information Service Algorithm Filing System, created after gig-workers went on strike in 2022 because algorithms in China’s two largest food delivery apps subjected workers to grueling delivery schedules, seems to demand more detail as it forces companies to come to discuss their design choices in short-response answers \cite{CEIP22}. With sections like “Algorithm Risk and Prevention Mechanism” and “Algorithm security self-assessment”, one would expect a productive discussion connecting technical design choices with real societal outcomes (or at least the security and stability outcomes that the Chinese government cares about). While disclosing technical details might seem self-explanatory, the critical technical details and correlating social outcomes are unclear. For example, while it is possible to speak at a high level about how systems bring about social outcomes, it is not apparent how interactions between specific system components may contribute to social unrest, even though technical specificity is needed to be able to reflect and causally explain particular impacts. This means companies must wrestle with a seemingly inescapable dilemma and may default to explaining their systems with  “a mix of metaphor and simplified language”. When Carnegie Endowment for International Peace scholars Matt Sheehan and Sharon Du examined a partially-disclosed report on Weibo’s “hot search” algorithm, they concluded that the information included was “so high-level that an observer with no knowledge of this specific algorithm could guess it” or anticipate its effects ~\cite{CEIP222}. 

To sum up, algorithm impact assessments in Canada and China make the opposite mistake as Model Cards: they recognize the forest, but (for different reasons) neglect the trees. None prioritize the feedback-laden nature of deployed ML systems by ensuring that designers update and continuously monitor the effects of their systems.

\subsection{Reward Reports as an Interface for Dynamic Documentation}

An ideal AI documentation system should create a framework that allows for deliberation between the forest and trees, matching high-level performance outcomes with specific, granular design decisions. While AI seldom solely utilizes reinforcement learning, the language of reinforcement learning may prove to be the right metaphor to discuss the kind of iterative decision-making fueling deployed intelligent systems. When AI is not explicitly framed as reinforcement learning processes and incorporates data from the environment to some end set by designers, it is useful to construe that system’s actions over time as reinforcement learning. When understood in these terms, AI documentation should revolve around examining the evaluative criteria and objectives, or the rewards, set by designers and the ways in which feedback is collected and utilized. Documenting the consequences of algorithmic changes in such systems would not only help regulators propose targeted and constructive guidance, but it would also shed light on the constellation of factors–both within the technical design of the system and in its environment–that underlie decisions made by otherwise opaque models. Such a tool would permit designers of AI systems to publicly and continuously assess their work over time, piecing together users’ comments and crowd-sourced data to form a picture of how their system is actually interacting with the world. 

One example of a documentation system that attempts to study how AI interacts with its environment post-deployment is Reward Reports: a form with prompts forcing AI designers to detail their motivations and expectations for their system and continuously assess the system’s performance post-deployment \cite{gilbert2022reward}. Reward Reports are composed of six sections. The language of reinforcement learning emerges when discussing the second section, Optimization Intent, which inquires about the performance metrics and failure modes, and the third section, Institutional Interface, which asks which external entities the system will engage with and how the system will remain accountable to these stakeholders. A Reward Report will also require engineers to disclose design elements critical to how data is processed and how user feedback is translated into performance metrics and fine-tune the system in the Implementation and Evaluation sections.  Lastly, the System Maintenance section contains a changelog where designers log any changes made to the ML system \cite{gilbert2022reward}.

\section{The Components of Dynamic Documentation: Feedback, Metrics, Interfaces}

Nevertheless, Reward Reports falls short of being able to track system performance on key metrics over time. The goal of a documentation system for AI should be to take continuous snapshots of a system’s performance, in a similar way to how Model Cards can provide a rich picture of a single model’s biases and behavior at a single point in time. This can be achieved if Reward Reports interacted dynamically with user interactions. However, the ways in which users interact with ML systems vary significantly, and it may not be obvious how one should translate user feedback into a Reward Report in every situation. Thus, Reward Reports would also need to utilize a standardized benchmark to process and accommodate diverse modes of user interaction.

For LLMs, systems like the Holistic Evaluation of Language Models (HELM) have already successfully laid out metrics that assess performance based on real interactions with the model \cite{liang2022holistic}.  HELM’s framers are cognizant of the challenge of comparing the performance of different language models with varied levels of accessibility that are geared toward different purposes and trained on distinct datasets. Thus, they use targeted examinations placing language models in the same scenarios under standardized conditions in which they are especially vulnerable to producing inappropriate generations, using the same few-shot prompting for all models. Each model is assessed on seven metrics: accuracy, calibration, robustness, fairness, bias, toxicity, and efficiency. 

In many ways, the HELM approach is analogous to the rationale behind Superfund site categorization. After a series of site inspections, the EPA uses a standardized metric 
called the Hazard Ranking System to ascertain the relative risk sites pose to environmental or human health. Just as HELM facilitates comparison amongst language models, the Hazard Ranking System allows the EPA to compare and evaluate a range of disasters in a variety of contexts. Based on the uniform threshold of a hazard risk score, sites are evaluated as eligible for long-term EPA intervention and placement on the National Priorities List 
\cite{holifield2012environmental,EPA23}. Just as the Hazard Risk Score flattens a site-specific evaluation into a single rank, a series of metrics like HELM can reduce model-specific user comments into something that a general documentation system like Reward Reports can handle. Continuous user feedback, treated like public comments in EISs, consumer complaints held by the Consumer Financial Protection Bureau \cite{CFPB23}, or the Federal Trade Commission’s Sentinel database \cite{FTC23} could be evaluated according to metrics like HELM and used to inform regular iterations of Reward Reports. Therefore, the key to making a dynamic documentation system to AI tractable is a user interface that can collect and categorize user feedback along metrics like HELM, which can then be reflected on and formatted onto a framework like Reward Reports.

\section{Conclusion}

Dynamic documentation remains an open and pressing problem. In this paper, we have sketched the critical design problems that make this approach necessary for AI systems (vs. algorithms and automated decision mechanisms). The limitations of past, present, and recently proposed documentation protocols provide valuable lessons for how fully dynamic documentation would need to be structured in order for the behavior of increasingly capable systems to be understood by their designers, let alone evaluated by key stakeholders.


\bibliographystyle{ACM-Reference-Format}
\bibliography{sample-acmlarge}



\end{document}